\documentclass[twocolumn,showpacs,preprintnumbers,amsmath,amssymb,superscriptaddress]{revtex4}
\usepackage[dvips]{graphicx}
\usepackage{dcolumn}
\usepackage{bm}
\usepackage[titletoc,title]{appendix}
\usepackage{longtable}
\usepackage{bm}
\usepackage{color}

\begin{document}
\setlength\LTcapwidth{6.5in}
\setlength\LTleft{0pt}
\setlength\LTright{0pt}














\title{Critical point determination from probability distribution functions in the three dimensional Ising model}

\author{Francisco Sastre}
\address{Departamento de Ingenier\'ia F\'isica, Divisi\'on de Ciencias e
Ingenier\'ias, Campus Le\'on de la Universidad de Guanajuato, AP E-143, CP 37150,
Le\'on, Guanajuato, M\'exico}

\begin{abstract}
In this work we propose a new numerical method to evaluate the critical point, the susceptibility critical
exponent 
and the correlation length critical exponent of the three
dimensional Ising model without external field using an algorithm that evaluates directly
the derivative of the logarithm of the probability distribution function with respect to the
magnetisation. Using standard finite-size scaling theory we found that correction-to-scaling effects
are not present within this approach.
Our results are in good agreement with previous reported values for the three dimensional Ising model.
\end{abstract}




\maketitle

\section{Introduction}

The Ising model has a great importance in statistical mechanics since a great variety of techniques
and methods, analytical and numerical, have been formulated first on this model.
There are several numerical algorithms that can be used to study
the critical behavior in spin systems, we can mention three types:
\begin{itemize}
\item Those that require adjusting a
control parameter, like the standard Monte Carlo or the Wolff algorithms~\cite{Wolff1989}.
\item Algorithms that not required to fine tune any parameter, examples of those kind are the Invasion Cluster
Algorithm~\cite{Machta1995}, algorithms based on Self Organisation~\cite{Fulco1999} or
the Locally Cluster Algorithm~\cite{Faraggi2008}. 
\item Algorithms that evaluate the Density of States
in the microcanonical ensemble~\cite{Wang2001,Huller2002}.
\end{itemize}

In this work we propose a new methodology based on the algorithm 
proposed by Sastre {\it et al.} for the evaluation of effective temperatures in out of equilibrium Ising-like systems~\cite{Sastre2003}. 
The algorithm is a canonical generalisation of the work 
proposed by H\"uller and Pleimling~\cite{Huller2002}
for the evaluation of the density of states in the two and three dimensional Ising model. It is important to point out that variations of this algorithm have been successfully implemented
in fluids with discrete potential interaction. The microcanonical version was used in~\cite{Sastre2015} to evaluate thermodynamic properties in the supercritical region and in~\cite{Sastre2018}
for the evaluation of the Hight Temperature Expansion coefficients of the Helmholtz free energy. The canonical version~\cite{Sastre2020} was used for the evaluation of the critical temperature
and the correlation length critical exponent in the Square-Well fluid with interaction range of $1.5$ times the particle diameter.

Our aim in this work is to prove that a completely new methodology can be used to study critical phenomena on Ising-like systems.
In particular we want to evaluate the critical temperature and the critical
exponents for the correlation length, $\nu$, and the susceptibility, $\gamma$, in the three dimensional Ising model in an efficient way.

This article is organized as follows: in section~\ref{seccionprobas} we explain the basic definitions
for the Ising model and the algorithm used in this work, in section~\ref{seccioncritica} we apply the
method to evaluate the critical point and the critical exponents.  We made our concluding remarks in
the section~\ref{conclusiones}.

\section{Probability distributions for the Ising model}
\label{seccionprobas}

For a better understanding of the algorithm used in this work we will
review first how the algorithm proposed by H\"uller and Pleimling in the
microcanonical ensemble works. The algorithm uses a variation of the transition variable
method~\cite{Oliveira1996,Oliveira1998,Kastner2000} in order to evaluate the
entropy as function of the energy and the magnetisation.

The hamiltonian for the Ising model on a three dimensional cubic lattice without external field and nearest neighbors interaction is 
\begin{equation}
\frac{1}{k_B T} H=-\beta^{-1}\sum_{\langle i,j\rangle} \sigma_i \sigma_j,
\end{equation}
where $\sigma_i = \pm 1$ is the spin in the $i$~th site, $\beta=k_B T/J$  is the control parameter and $J>0$ is the coupling
between first nearest neighbor
spins. The notation $\langle i,j \rangle$ indicates that the summation runs over all nearest neighbors pairs 
on the lattice. If we consider a system with periodic boundary conditions and $N=L^3$ spins, the magnetisation, $M=\sum_i \sigma_i$, and the energy
will be bounded. Moreover, when a spin is flipped in the system we observe that 
$\Delta M = \pm 2$ and $\Delta E/J = \pm 4\eta$, $\eta = 0,~\pm 1,~\pm 2,~\pm 3$.

In the standard microcanonical notation $\Omega(E_{\mu},M_k)$ is the number of microstates
that share the same magnetisation $M_k$ and energy $E_{\mu}$, for simplicity we will use
$\Omega_{\mu,k}$ for this number. When the systems is in a given macrostate $\Omega_{\mu,k}$ and we
flip a spin at random
we can reach a new macrostate $\Omega_{\nu,l}$, with $E_\nu = E_{\mu}+4\eta J$ and $M_k = M_l \pm 2$. The probability of reach $\Omega_{\nu,l}$ starting from  $\Omega_{\mu,k}$ will be given by
\begin{equation}
P^{(m)}_{(\mu,k),(\nu,l)} = \frac{V_{(\mu,k),(\nu,l)}}{N\Omega_{\mu,k}}, \label{adelante}
\end{equation}
the superscript $(m)$ denotes that we are working in the microcanonical ensemble and $V_{(\mu,k),(\nu,l)}$ indicates how many ways the system can reach $\Omega_{\nu,l}$ starting from $\Omega_{\mu,k}$, this quantity is purely geometric.
We can also obtain the reverse probability with
\begin{equation}
P^{(m)}_{(\nu,l),(\mu,k)} = \frac{V_{(\nu,l),(\mu,k)}}{N\Omega_{\nu,l}}. \label{inverso}
\end{equation}
As any spin flip can be reversed, the relation
$V_{(\mu,k),(\nu,l)}=V_{(\nu,l),(\mu,k)}$ must be satisfied. For example, from the base state with 
$E_\mu=-3NJ$
and $M_k=N$ we can reach the state with $E_\nu=-3J(N-4)$ and $M_l=N-2$ in $N$ ways,
then $V_{(\mu,k),(\nu,l)}=N$ and, as $\Omega_{\mu,k}=1$, $P^{(m)}_{(\mu,k),(\nu,l)}=1$. In the reverse process we have
$\Omega_{\nu,l}=N$ then $P^{(m)}_{(\nu,l),(\mu,k)}=1/N$. Combining equations (\ref{adelante}) and (\ref{inverso}) we obtain the
important microcanonical relation
\begin{equation}
\frac{P^{(m)}_{(\mu,k),(\nu,l)}}{P^{(m)}_{(\nu,l),(\mu,k)}}=\frac{\Omega_{\nu,l}}{\Omega_{\mu,k}}.\label{razones}
\end{equation}
The last equation can be used to evaluate the microcanonical derivatives
\begin{equation}
\frac{1}{T}=\frac{\partial S}{\partial E},
\end{equation}
 and
\begin{equation}
-\frac{B}{T} =\frac{\partial S}{\partial M},
\end{equation}
where $B$ is the magnetic external field.
The change on the entropy can be obtained using the following approximation
\begin{equation}
\Delta S = k_B \ln\left( \frac{P^{(m)}_{(\mu,k),(\nu,l)}}{P^{(m)}_{(\nu,l),(\mu,k)}}\right) \approx \Delta E \frac{1}{T} - \Delta M \frac{B}{T}.
\label{entropia}
\end{equation}
In the simulation the probabilities can be estimated with the rate of attempts $T_{(\mu,k),(\nu,l)}$
to go from $\Omega_{\mu,k}$ to $\Omega_{\nu,l}$. The rate is given by the relation
\begin{equation}
T_{(\mu,k),(\nu,l)}=\frac{z_{(\mu,k),(\nu,l)}}{z_{(\mu,k)}},
\end{equation}
where $z_{(\mu,k)}$ is the number of times that the system spends in a macrostate $\Omega_{\mu,k}$, 
and $z_{(\mu,k),(\nu,l)}$ is
the number of times that the system attempts to change from $\Omega_{\mu,k}$ to $\Omega_{\nu,l}$.
In this method, once that we fix the ranges $[E_{min},E_{max}]$ and $[M_{min},M_{max}]$, where the
simulation will be confined,
the quantities  $z_{(\mu,k),(\nu,l)}$ and $z_{(\mu,k)}$ can be estimated
in the following way:
\begin{enumerate}
\item With $E_{min} \le E_\mu \le E_{max}$  and $M_{min} \le M_k \le M_{max}$ as initial state, a spin is chosen at random and $z_{(\mu,k)}$ is always incremented by 1.
\item We evaluate the new values $E_\nu$ and $M_l$ that the system would take if the chosen spin is flipped.
\item If $E_{min} \le E_\nu \le E_{max}$ and $M_{min} \le M_l \le M_{max}$ the quantity $z_{(\mu,k),(\nu,l)}$ is incremented by 1.
\item The spin flip attempt is accepted with probability
$\min{(1,\frac{z_{(\nu,l),(\mu,k)} z_{(\mu,k)}}{z_{(\mu,k),(\nu,l)} z_{(\nu,l)}})}$. This
condition assures that all macrostates are visited with equal probability, independently
of their degeneracy.
\end{enumerate}
The values $z_{(\mu,k),(\nu,l)}$ and $z_{(\mu,k)}$ can be initialized with any positive integer,
a safe option is $1$, and after a large number of spin flip attempts we will observe that 
$T_{(\mu,k),(\nu,l)} \to P_{(\mu,k),(\nu,l)}$.

M\"uller an Pleimling used this method for the determination of the microcanonically defined spontaneous magnetisation and the order parameter critical exponent for the Ising model in two and three
dimensions. 
This algorithm is highly efficient for evaluate the ratios $\Omega_{\nu,l}/\Omega_{\mu,k}$ since
it gives the freedom of restricting the
calculations to a chosen range in the energy and magnetisation. Additional details of the method can be found in the
original work~\cite{Huller2002}.

The microcanonical algorithm counts all the attempts to change from a given macrostate to another, as long as the final macrostate is an allowed one,
while the generalisation proposed in~\cite{Sastre2003}
adds an additional condition to the attempts count. The additional condition includes a "heat bath", then we will need to incorporate an extra factor in the ratio of probabilities
\begin{equation}
\frac{P_{(\mu,k),(\nu,l)}}{P_{(\nu,l),(\mu,k)}}=
\frac{P^{(m)}_{(\mu,k),(\nu,l)}}{P^{(m)}_{(\nu,l),(\mu,k)}} e^{-\frac{\Delta E}{k_B T}}=\frac{\Omega_{\nu,l}}{\Omega_{\mu,k}},
e^{-\frac{\Delta E}{k_B T}}
\label{canonical}
\end{equation}
here the absence of the superscript indicates that we are no longer in the microcanonical ensemble.
Combining Equations (\ref{entropia}) and (\ref{canonical}) we get
\begin{equation}
\ln({P_{(\mu,k),(\nu,l)}})-\ln({P_{(\nu,l),(\mu,k)}}) \approx -\Delta M \frac{B}{k_B T},
\end{equation}
where we can drop the subscript $\mu$ and $\nu$, since the right hand side of the equation depends only on $\Delta M$.
In the simulation the probabilities now can be estimated with the rate of attempts $T_{kl}$ to go from a macrostate with magnetisation
$M_k$ (level $k$) to a macrostate with magnetisation $M_l$ (level $l$). The quantity $T_{kl}$ will be given now by the relation
\begin{equation}
T_{kl} = \frac{z_{kl}}{z_{k}},
\end{equation}
where $z_{kl}$ is the number of times that the system attempts
to change from level $k$ to level $l$ and $z_k$ is the number of times that the system spends in level $k$. For the estimation of the $z_{kl}$ and $z_{k}$ values, the detailed steps are now:
\begin{enumerate}
\item With $M_{min} \le M_k \le M_{max}$ as initial state, a spin is chosen at random and $z_{k}$ is always incremented by 1.
\item If the possible state $M_l$, that would be reached if the chosen spin is flipped, is allowed we evaluate $\Delta E$ between 
states $M_k$ and $M_l$, and the quantity $z_{kl}$ is incremented by 1 with probability $\min(1,e^{-\Delta E/k_B T})$.
\item The spin flip attempt is accepted with probability
$\min{(1,\frac{z_{lk} z_{k}}{z_{kl} z_{l}})}$.
\end{enumerate}
The values $z_{kl}$ and $z_{k}$ are initialized to 1
and after a large number of spin flip attempts we will observe that 
$T_{kl} \to P_{kl}$.

Now we obtain the derivative of $\ln{P}$, instead of the derivatives of $S$, with the following approach
\begin{equation}
\left.\frac{\partial \ln{P}}{\partial M}\right|_k \approx \ln{P_{kl}}-\ln{P_{lk}}  \approx \ln(T_{kl}/T_{lk}).
\end{equation}
We will use the following function in our simulations
\begin{equation} 
g(m)=L^{-d}\frac{\partial \ln P}{\partial m}, \label{canonical2}
\end{equation}
where $m=M/L^d$, since it can be obtained directly from the transition rates with the approximation
\begin{equation}
g(m_k)\approx \frac{1}{2}\left(\ln[(T_{k,k+1})/(T_{k+1,k})]-\ln[(T_{k,k-1})/(T_{k-1,k})]\right),
\label{bonequation}
\end{equation}
where $T_{k,k\pm k}$ are the transition rates from level $k$ to its adjacent levels and $T_{k\pm 1,k}$ are the
transition rates from the adjacent levels to level $k$.

\begin{figure}[ht]
\begin{center}
\includegraphics[width= 10.0cm]{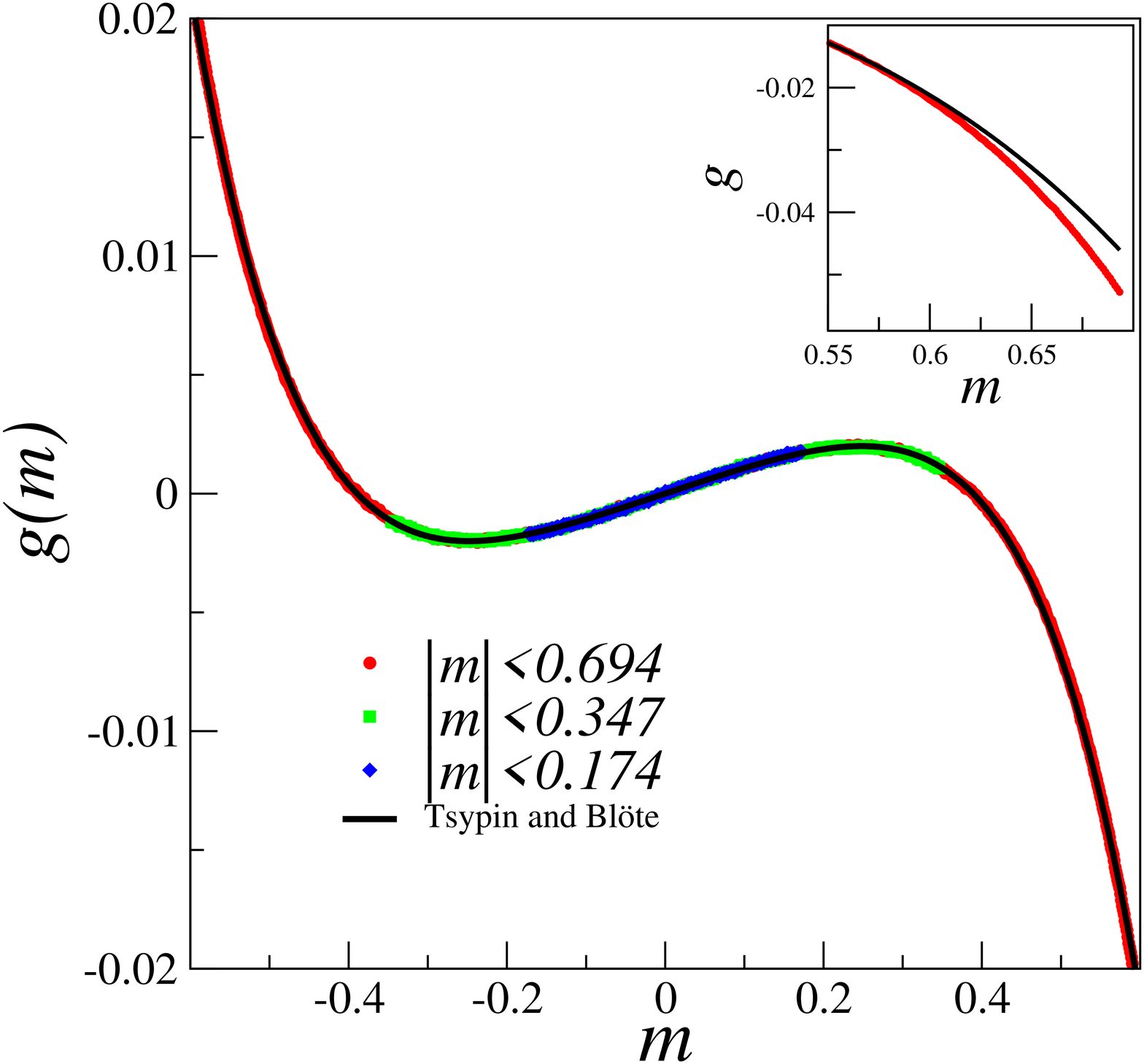}
\caption{\label{preuve} (Color online) Graph of $g$ as function of $m$ for three different range simulations.
We observe that the restriction in range does not affects the computed value of $g$. The continuous line
reproduces Eq.~(\ref{ansatzprob}) with the parameters $a=0.268,~c=0.859$ and $m_0=0.3892$
for $L=12$. There is a really good agreement with our results,
except in the extremes of the curves, as shown in the inset.
}
\end{center}
\end{figure}

We verified that
this method is compatible with the ansatz proposed by Tsypin and Bl\"ote\cite{Tsypin2000} 
for the three dimensional Ising model at the critical point
\begin{equation}
P(m)\sim \exp{\left[-\left(\left(\frac{m}{m_0}\right)^2-1\right)^2\left(a\left(\frac{m}{m_0}\right)^2+c\right)\right]},
\label{ansatzprob}
\end{equation}
where $a,~c$ and $m_0$ are size depending fitting parameters. For $L=12$ we performed three independent simulations
at $\beta_c= 0.221654$, taken from Ref.~\cite{Campbell2011},
with different ranges in $m$. Our results are show in Figure~\ref{preuve} along with the curve given
by Eq.~(\ref{ansatzprob}), using $a=0.268$, $c=0.268$, $m=0.3892$. Our simulations are in
really good agreement with the ansatz, except at the extreme values of
the curves, but this effect is also present in the results published by Tsypin and Bl\"othe.

\section{Results}
\label{seccioncritica}

\begin{figure}
\begin{center}
\includegraphics[width= 10.0cm]{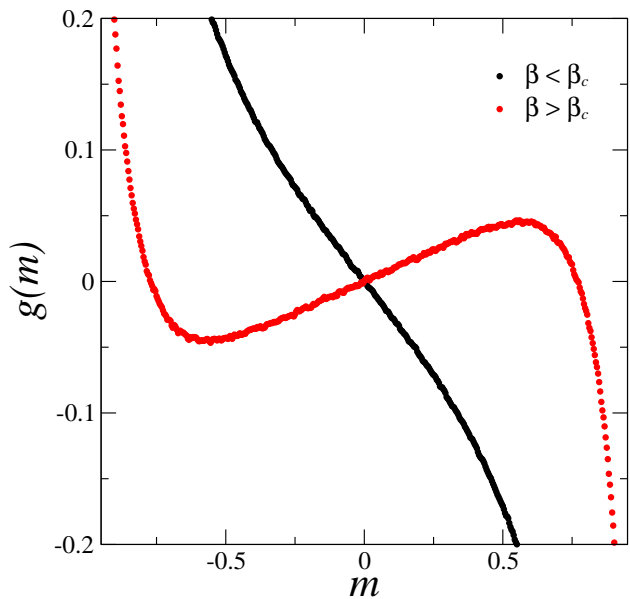}
\caption{\label{rompimiento} (Color online) Symmetry breaking for 
the three dimensional Ising model with $L=8$. For $\beta<\beta_c$ the curve cross
the horizontal axis at $m=0$ with a negative slope. For $\beta>\beta_c$ we have three crossings, at $m=0$
and at $m=\pm m_{sta}$  and we observe that the slope at $m=0$ is now positive.
}
\end{center}
\end{figure}

Now we will explain how to obtain the critical temperature and the correlation length critical exponent.
It is a well known fact that there is a change in the probability distribution of the order parameter above
and below a certain value of $\beta$, denoted as $\beta_c(L)$, that depends on the system size. For $\beta$ values
below $\beta_c(L)$ there is just one peek at $m=0$, that means that the curve $g(m)$ crosses the horizontal axis with
a negative slope. For $\beta$ values above $\beta_c(L)$ there are two peeks at $m=\pm m_{sta}$ in the probability distribution, while $g(m)$
crosses the horizontal axis in three places, at  $m=\pm m_{sta}$ with a negative
slope and at $m=0$ with a positive slope.
In Figure~\ref{rompimiento} we can see clearly the symmetry breaking for three dimensional Ising model
with linear size $L=8$.

\begin{figure}
\begin{center}
\includegraphics[width=10.0cm]{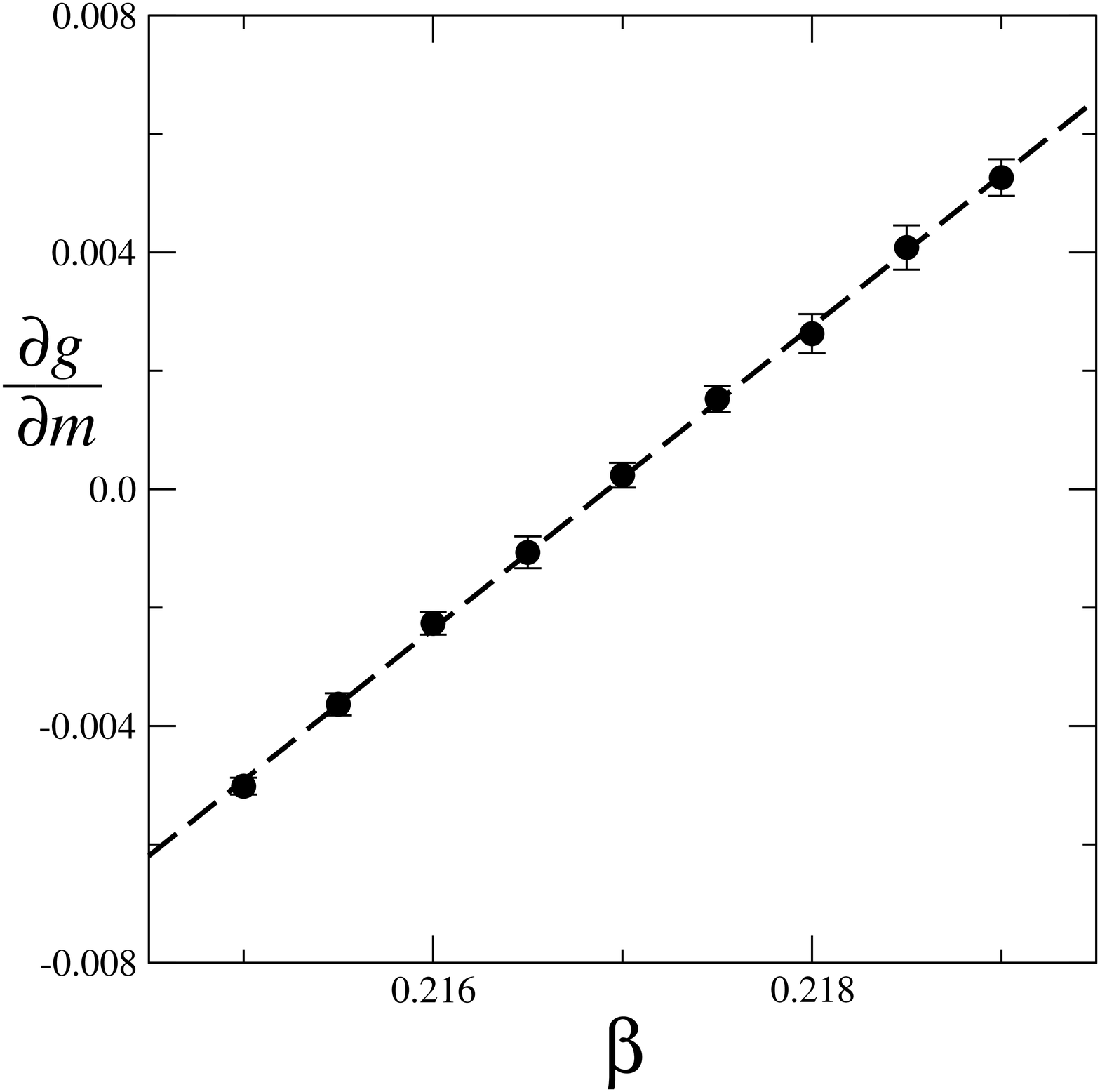}
\caption{\label{slopes} Slopes of the function $g(m)$ at $m=0$ for the three dimensional Ising model
with $L=12$ as function of the control parameter $\beta$. The dashed line is a linear fit to the simulation data. 
}
\end{center}
\end{figure}

As we can see there is a change in the sign of the slope $\partial g/\partial m$ around $m=0$ as function of $\beta$, then
we can find $\beta_c(L)$ restricting our simulations around $m=0$.
We performed simulations restricting the intervals to $|m| \lesssim 0.1$ on several system sizes and temperature values in
the three dimensional Ising model.
The justification for this interval is that $g(m)$ is linear around $m=0$  and the slope can be easily obtained from a linear fit
to the simulation data.
For every system size we evaluate the slope for several values of $\beta$, in Figure~\ref{slopes} we
are illustrating how the evaluation of $\beta_c(L)$ is performed for the case
$L=12$. Here we used a linear fit to the curves of the slope as function of $\beta$ in order to solve
$\left(\frac{\partial g}{\partial m}\right)_{\beta=\beta_c} = 0 $.

From here we can use the Finite size scaling ansatz for the critical temperature, given by the
relation
\begin{equation}
\beta_c(L)\approx \beta_c + A L^{-1/\nu},\label{scaling_temp}
\end{equation}
where $\beta_c$ is the critical control parameter for the infinite system, $A$ is a non universal parameter
and $\nu$ is the critical exponent for the correlation length~\cite{Ferrenberg1991}.
In principle Eq.~(\ref{scaling_temp}) is valid for sufficiently large $L$ values. For small systems the
last term changes to $L^{-1/\nu} (1+B L^{-\omega})$, here the parameter $\omega$ is the scale correction exponent,
whose reported value for the three dimensional Ising model is $\omega\approx 0.81$~\cite{Lundow2010}.

\begin{figure}
\vspace*{10pt}
\begin{center}
\includegraphics[width=10.0cm,clip]{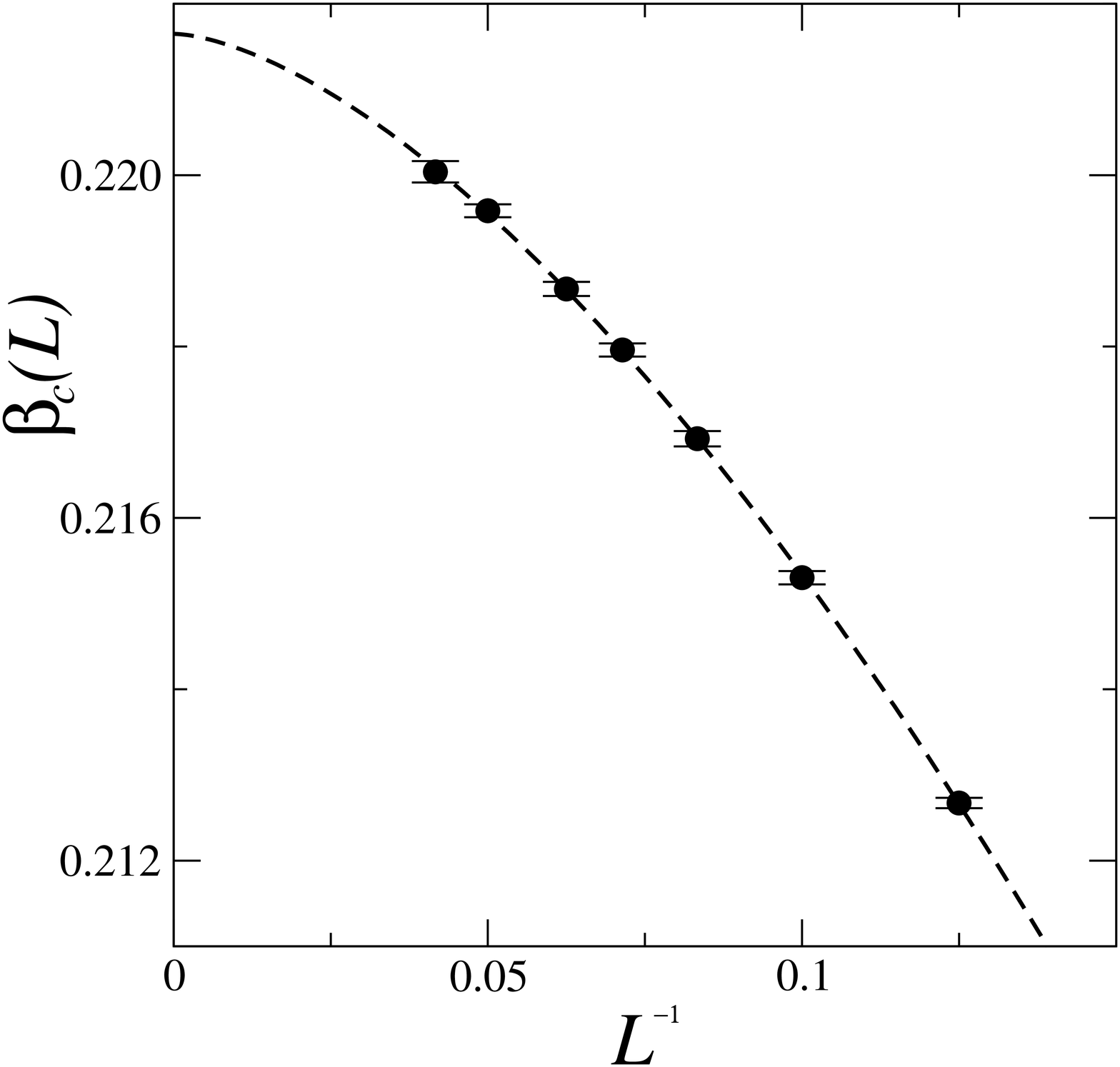}
\end{center}
\caption{\label{critico} Evaluation of the critical temperature and the correlation length critical exponent
for the three dimensional Ising model. The dashed line is a non-linear curve fit  to Eq. (\ref{scaling_temp}). The results from the
fit are 
$\beta_c=0.22165$, $A=-0.2439$ and $1/\nu=1.587$.
}
\end{figure}

The simulations were carried out in systems with linear sizes $L=8$, 10, 12, 14, 16, 20 and 24, using $5N\times 10^6$
spin flip attempts and 120 independent runs for every set of parameters.
With these values we obtain reliable data with less CPU time compared with standard canonical simulations, since most of
the spin flips are discarded when the system falls outside of the restricted range. However, all attempts,
successful or not, are used for the evaluation of the transition rates.
We evaluated $T_c$ and $\nu$ performing a non-linear curve fitting to Eq.~(\ref{scaling_temp}),
in Figure~\ref{critico} we show the evaluation of the critical point along with the $\nu$ critical exponent.
We must emphasized that in our analysis the scale correction exponent is absent, which is a great advantage in numerical simulations that study critical phenomena. This feature is also observed in the evaluation of the critical temperature for the square-well fluid using an equivalent method~\cite{Sastre2020}.
We think that the absence of scale corrections are related to the fact that in this method we evaluate the critical point analyzing the behavior of the
 probability distribution of the order parameter around $M=0$, while in most traditional methods the critical point is evaluated analyzing the behavior around the peaks of the probability distribution. 

The results for the critical point is
$\beta_c = 0.22165(65)$ and for the correlation length critical exponent we obtain $\nu = 0.6301(88)$, that are in good agreement with previous reported values, see Table~\ref{happyend}.
\begin{center}
\begin{table}[ht!]
\caption{\label{happyend}
Critical parameters for the three dimensional Ising model obtained in this work and those from literature.
The values between parenthesis indicate the uncertainty in the lasts digits.
}
\begin{tabular}{ccccc}
\hline
\hline
$\beta_c$ &  ~ & $\nu$ & ~ & Source\\
\hline
  0.22165(65)   & ~ & 0.6301(88)  & ~ &  This work \\
  0.22165452(8) & ~ & 0.63020(12) & ~ & Butera and Comi~\cite{Butera2002} \\
  0.221655(2)   & ~ & 0.6299(2)   & ~ & Deng and Bl\"ote~\cite{Deng2003} \\
  0.221654(2)   & ~ & 0.6308(4)   & ~ & Lundow and Campbell~\cite{Lundow2010} \\
\hline
\hline
\end{tabular}

\end{table}
\end{center}

Once that we have the critical temperature we can proceed to evaluate
the susceptibility critical exponent $\gamma$. We used the scaling
ansatz for the probability distribution function at the critical point
proposed in Ref.~\cite{Kaski1984}

 \begin{equation}
    P(m,L)\approx \exp{(-A_0+A_2 x^2+A_4 x^4+\dots)},
    \label{ansatzprob1}
    \end{equation}
where $x=mL^{\beta/\nu}<<1$ and $\beta$ is the order parameter critical exponent.
As we are restricting our simulations to the range $|m|\le0.1$ we can use the next approximation
 \begin{equation}
    \frac{\partial \ln{P}}{\partial m} = 2 A_2 m L^{2\beta/\nu}+O(x^3),
  \label{scalingprob1}
  \end{equation}
that we can combine with Eq.~(\ref{canonical2}) to obtain the desired scaling relation
 \begin{equation}
    \frac{\partial g}{\partial m} \sim L^{-\gamma/\nu}.
  \label{scalingprob}
  \end{equation}

\begin{figure}
\vspace*{10pt}
\begin{center}
\includegraphics[width=10.0cm,clip]{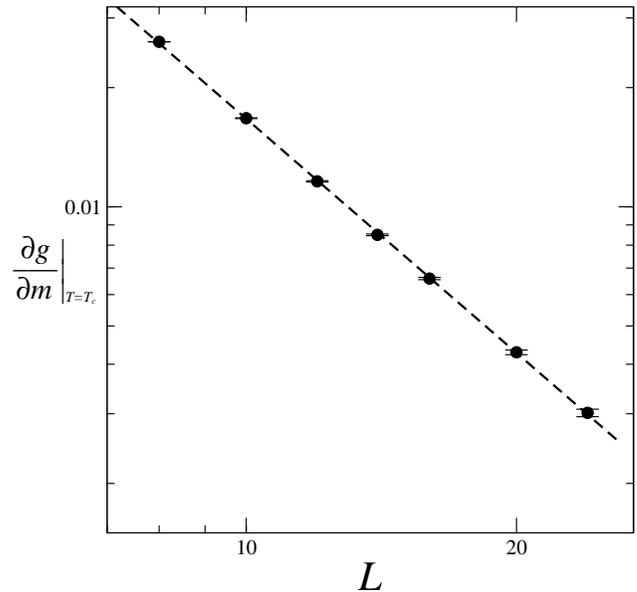}
\end{center}
\caption{\label{gamma3d} Evaluation of the critical exponent $\gamma/\nu$ for 
three dimensional Ising model. We are showing
a log-log graph of the
derivatives $\partial g/\partial m$ as function of the linear size $L$. The dashed line is
the linear fit to Equation (\ref{scalingprob}). From the fit we obtain the slope $\gamma/\nu=1.973$.
}
\end{figure}

For the three dimensional case we performed simulations at our estimated critical point $\beta_c=0.22165$
and linear sizes
$L=8$, 10, 12, 14, 16, 20 and 24. In this case we used $N\times10^7$ spin flip attempts and 120 independent runs for every $L$ value.
The critical exponent $\gamma$ was obtained from a linear fit to Eq.~(\ref{scalingprob}) as shown
in Figure~\ref{gamma3d}. From this fit 
we obtain
$\gamma/\nu=1.973(10)$ that is also in good agreement with the
reported value $\gamma/\nu=1.9632(6)$~\cite{Campbell2011}.
Again we observe that scaling correction exponents are absent.

\section{Conclusions}
\label{conclusiones}
We have presented a new method for the evaluation of the critical temperature and the critical exponents
for the correlation length $\nu$ and the susceptibility $\gamma$ on the three dimensional Ising model. Using the derivatives of the
probability distribution function for the magnetisation and small system sizes we obtain reliable results that are in good agreement with
the reported values. The method can be used
in a restricted range
on the magnetisation and this feature reduces the computational time in the simulations. 
One additional advantage of the method is that scale corrections are not present, at least in the three dimensional Ising model. In future works we will study if this advantage is present in other systems.

\section{Acknowledgements}
This research was
supported by Universidad de Guanajuato (M\'exico) under Proyecto DAIP 879/2016 and CONACyT (M\'exico)(grant CB-2017-2018-A1-S-30736-F-2164).

\section*{References}

\bibliography{referencias}

\end{document}